\documentclass[sigconf,screen]{acmart}

\AtBeginDocument{%
  }

\copyrightyear{2024}
\acmYear{2024}
\setcopyright{rightsretained}
\acmConference[ASSETS '24]{The 26th International ACM SIGACCESS Conference on Computers and Accessibility}{October 27--30, 2024}{St. John's, NL, Canada}
\acmBooktitle{The 26th International ACM SIGACCESS Conference on Computers and Accessibility (ASSETS '24), October 27--30, 2024, St. John's, NL, Canada}
\acmDOI{10.1145/3663548.3688514}
\acmISBN{979-8-4007-0677-6/24/10}

\usepackage{listings}
\usepackage{xcolor}

\colorlet{punct}{red!60!black}
\definecolor{background}{HTML}{F7F7F7}
\definecolor{delim}{RGB}{20,105,176}
\colorlet{numb}{magenta!60!black}

\lstdefinelanguage{json}{
    basicstyle=\normalfont\ttfamily,
    numbers=left,
    numberstyle=\scriptsize,
    stepnumber=1,
    numbersep=8pt,
    showstringspaces=false,
    breaklines=true,
    frame=lines,
    backgroundcolor=\color{background},
    literate=
     *{0}{{{\color{numb}0}}}{1}
      {1}{{{\color{numb}1}}}{1}
      {2}{{{\color{numb}2}}}{1}
      {3}{{{\color{numb}3}}}{1}
      {4}{{{\color{numb}4}}}{1}
      {5}{{{\color{numb}5}}}{1}
      {6}{{{\color{numb}6}}}{1}
      {7}{{{\color{numb}7}}}{1}
      {8}{{{\color{numb}8}}}{1}
      {9}{{{\color{numb}9}}}{1}
      {:}{{{\color{punct}{:}}}}{1}
      {,}{{{\color{punct}{,}}}}{1}
      {\{}{{{\color{delim}{\{}}}}{1}
      {\}}{{{\color{delim}{\}}}}}{1}
      {[}{{{\color{delim}{[}}}}{1}
      {]}{{{\color{delim}{]}}}}{1},
}

\usepackage{url}
\usepackage{hyperref}
\usepackage[hyphenbreaks]{breakurl}

\begin{document}

\title{Speech-based Mark for Data Sonification}

\author{Yichun Zhao}
\email{yichunzhao@uvic.ca}
\affiliation{%
  \institution{University of Victoria}
  \city{Victoria, BC}
  \country{Canada}
}

\author{Jingyi Lu}
\email{jingyilu@uvic.ca}
\affiliation{%
  \institution{University of Victoria}
  \city{Victoria, BC}
  \country{Canada}
}

\author{Miguel A. Nacenta}
\email{nacenta@uvic.ca}
\affiliation{%
  \institution{University of Victoria}
  \city{Victoria, BC}
  \country{Canada}
}


\begin{abstract}
Sonification serves as a powerful tool for data accessibility, especially for people with vision loss. Among various modalities, speech is a familiar means of communication similar to the role of text in visualization. However, speech-based sonification is underexplored. We introduce \textit{SpeechTone}, a novel speech-based mark for data sonification and extension to the existing \textit{Erie} declarative grammar for sonification. It encodes data into speech attributes such as pitch, speed, voice and speech content. We demonstrate the efficacy of \textit{SpeechTone} through three examples. 
\end{abstract}




\maketitle

\section{Introduction}

Data accessibility is a critical issue in today's data-driven world. 
This is particularly true for audiences with vision loss, for whom traditional visual data representations might not be accessible. While alternative methods such as alt-text or data tables can provide some level of accessibility, they often lack the richness and intuitiveness of the original data representation (e.g., \cite{Chintalapati_Bragg_Wang_2022, Wang_Wang_Jung_Kim_2022}). In this context, sonification can serve as a powerful tool and a transformative technique that converts data into auditory experiences. 
Kim et al.~\cite{Kim_Kim_Hullman_2024} and Enge et al.~\cite{Enge_Rind_Iber_Höldrich_Aigner_2023} have applied the principles of information visualization to structure and improve the auditory representation of data, drawing an analogy between an auditory tone and a visual mark (which is a basic unit to represent data). 
However, the use of sonification still has its challenges~\cite{SonificationInclusiveLearningDesignHandbook, Neuhoff_2019}. While there are numerous tools available for sonification, they often require users to learn new ways of interpreting data. This learning curve can pose a barrier to accessibility, potentially limiting the reach and effectiveness of sonification.

To help overcome such challenges, we propose a speech-based mark for sonification. Speech holds significant potential as a method of sonification. It is a common modality that we use every day, making it a familiar and accessible means of communication. For individuals with vision loss, speech can serve as a natural and intuitive way to interact with data, especially for those who are familiar with screen readers. Speech has been used as a basic modality to convey labels or raw data in sonification, but we find opportunities in directly manipulating characteristics of speech to enhance the expressiveness of speech sonification. For instance, the \textit{JAWS} screen reader~\cite{JAWS} pronounces the accessible header of a spreadsheet column with a reduced pitch, a distinct tonal variation that helps users understand that it signifies a particular meaning. 
In our work, we leverage and extend the declarative grammar for sonification, \textit{Erie}~\cite{Kim_Kim_Hullman_2024}, and present a new speech-based mark for sonification, \textit{SpeechTone}, which enhances the computation of auditory data representations. Additionally, we provide demonstrations of \textit{SpeechTone}.

\section{Related Work}

Various sonification tools and toolkits exist~\cite{Benjamin_et_al_2007, Enge_et_al_2022, Frame, Lindetorp_Falkenberg_2021, Pauletto_Hunt, Phillips_Cabrera_2019, Reinsch_Hermann_2023, Walker_Cothran_2003, Worrall_Bylstra_Barrass_Dean_2007, Zhao_Tzanetakis_2022}, and a recent review~\cite{Kim_Kim_Hullman_2024} has provided a thorough walkthrough of the audio encoding and features these tools support. However, most of them do not support speech. The ones that support speech are usually developed for sonification in data visualizations such as charts~\cite{Alam_Islam_Hoque_2023, Sharif_Wang_Muongchan_Reinecke_Wobbrock_2022, Thompson_Martinez_Sarikaya_Cutrell_Lee_2023, Zong_Lee_Lundgard_Jang_Hajas_Satyanarayan_2022, Holloway_Goncu_Ilsar_Butler_Marriott_2022, Zong_PedrazaPineros_Chen_Hajas_Satyanarayan_2024, Fan_Siu_Law_Zhen_O’Modhrain_Follmer_2022}, 
maps~\cite{Zhao_Plaisant_Shneiderman_Lazar_2008, Kane_Morris_Perkins_Wigdor_Ladner_Wobbrock_2011, Poppinga_Magnusson_Pielot_Rassmus_Gröhn_2011, Kaklanis_Votis_Tzovaras_2013, Albouys_Perrois_Laviole_Briant_Brock_2018, Bahram_2013}, 
and other graphics and diagrams~\cite{Brown_Pettifer_Stevens_2003, Horstmann_Lorenz_Watkowski_Ioannidis_Herzog_King_Evans_Hagen_Schlieder_Burn_etal._2004, King_Blenkhorn_Crombie_Dijkstra_Evans_Wood_2004, Petrie_Schlieder_Blenkhorn_Evans_King_O’Neill_Ioannidis_Gallagher_Crombie_Mager_etal._2002, Petrie_King_Burn_Pavan_2006, Zhang_Wobbrock_2022, Zhao_Nacenta_Sukhai_Somanath_2024}. 
These tools provide speech as a method to communicate fundamental data by directly vocalizing it, or by articulating the associated labels. Yet, speech-based encoding remains unexplored.

\textit{Erie}~\cite{Kim_Kim_Hullman_2024} 
is a declarative programming grammar and a JavaScript library designed for data sonification. It allows users to directly express audio attribute mappings via specifications, supporting various editable channels for mapping individual data attributes with unique modifications. This approach simplifies the process of data sonification, making it more accessible to users who may not be familiar with sound synthesis methods. \textit{Erie} also supports data manipulation and transformation, a feature missed by previous sonification tools. We use the developmental framework of \textit{Erie} and extend it to include the \textit{SpeechTone} mark.

\section{\textit{SpeechTone} Design}

Data visualization commonly uses text as marks~\cite{Brath_2020} (e.g., infotypography~\cite{Lang_Nacenta_2022}, tables, and bar charts annotated with numbers) which serve as a key method for representing and communicating data visually. The design space of text allows for the representation of additional information via text labels in visualizations. 
Analogously, we design the speech-based mark, \textit{SpeechTone}, to leverage auditory channels to convey information instead of visual cues, providing a more direct layer of access to data. 

The design of \textit{SpeechTone} followed an iterative process. We considered various attributes and marks for speech-based sonification and provided an initial design. We tested it with a low-fidelity version using canonical datasets and iterated on the design, making necessary adjustments and improvements. 

We implemented the \textit{SpeechTone} mark by extending the \textit{Erie} grammar to include it as an alternative approach to data mapping using speech. Conceptually, \textit{SpeechTone} allows for auditory encoding through speech attributes such as pitch, loudness, duration, speed, voice or timbre and text content (to be read out). 

There are currently four possible speech channels implemented by modifying the \textit{Erie} library: ``SpeechTonePitch'', ``SpeechToneSpeed'', ``SpeechToneVoice'' and ``SpeechToneText'', which encode data into speech pitch, rate, voice and text content respectively (Refer to Section~\ref{sec:demo} for their specifications): 

\begin{itemize}
    \item ``SpeechTonePitch'': This channel adjusts the speech pitch. It can range between 0 (lowest) and 2 (highest), with 1 being the default pitch of the voice.
    \item ``SpeechToneSpeed'': This channel modulates the speech rate. It can range between 0.1 (lowest) and 10 (highest), with 1 being the default rate. 
    \item ``SpeechToneVoice'': This channel enables different voice timbres using the Web Speech API in \textit{Erie}. Conceptually, it can be further broken down into high-level categories such as gender and accents, but the voice selections depend on the platform where the API is used. Its range specifies the voice IDs unique to the platform. 
    \item ``SpeechToneText'': This channel specifies the text to be read out. If its value is a data attribute name (e.g., ``year'', ``age'', ``date'' in a dataset), then the data of the attribute is vocalized. If no attribute is matched, \textit{SpeechTone} will directly read the text specified in this channel. 
    
\end{itemize}

To use \textit{SpeechTone} as part of the extended \textit{Erie} grammar, we specify \textit{SpeechTone} by introducing it as a new type in the ``Tone'' JSON object and declaring new channels in the ``Encoding'' JSON object which specifies the encodings with examples provided in Section~\ref{sec:demo}. In \textit{Erie}, a tone in sonification is analogous to a mark in visualization.

\begin{lstlisting}[language=json, numbers=none]
"tone": {
    "continued": false,
    "type": "speechtone"
}
\end{lstlisting}

\section{Demonstrations}
\label{sec:demo}

To illustrate the efficacy of \textit{SpeechTone}, we employed the ``Cars'' dataset, a common example used in \textit{Vega}~\cite{vega_data}, and sonified it using diverse strategies. We selected this dataset also because it includes multiple types of data (nominal, numerical, time-based and text-based). The ``Cars'' dataset contains various attributes of cars, such as model name, miles per gallon, horsepower, and year of production (Refer to Table~\ref{tab:cars} for a tabular snippet of it). We present three demonstrations with both audio samples (in supplementary material) and code samples (in JSON format supported by \textit{Erie}), targeted at the encodings supported by \textit{SpeechTone}. (We used Chromium-based browsers such as Edge or Chrome to demonstrate; there exist differences in the voices supported by different platforms. ) 

\newcommand*\rot{\rotatebox{50}}
\begin{table*}[!h]
\caption{Snippet of the ``Cars'' dataset in tabular format.}
\label{tab:cars}
\Description{This table shows a snippet of a dataset of various car models. The columns represent different characteristics of the cars: Name, Miles_per_Gallon, Cylinders, Displacement, Horsepower, Weight_in_lbs, Acceleration, Year and Origin. For example, it lists the 'datsun 1200' from Japan, which was released in 1971, and it runs 35 miles per gallon, has 4 cylinders, has a displacement of 72, has a horsepower of 69, weighs 1613 lbs and has an acceleration of 18. The table shows 5 other rows of the same structure and is truncated with ellipses, indicating that there are more rows of data not shown to save space.  
}

\begin{center}
\small
\begin{tabular}{ c c c c c c c c c  } 

\rot{Name} & \rot{Miles\_per\_Gallon} & \rot{Cylinders} & \rot{Displacement} & \rot{Horsepower} & \rot{Weight\_in\_lbs} & \rot{Acceleration} & \rot{Year} & \rot{Origin} \\ 
\hline
... &&&&&&&& \\ 

datsun 1200 & 35 & 4 & 72 & 69 & 1613 & 18 & 1971 & Japan \\ 

volkswagen model 111 & 27 & 4 & 97 & 60 & 1834 & 19 & 1971 & Europe \\ 

plymouth cricket & 26 & 4 & 91 & 70 & 1955 & 20.5 & 1971 & USA \\ 

toyota corona hardtop & 24 & 4 & 113 & 95 & 2278 & 15.5 & 1972 & Japan \\ 

dodge colt hardtop & 25 & 4 & 97.5 & 80 & 2126 & 17 & 1972 & USA \\ 

volkswagen type 3 & 23 & 4 & 97 & 54 & 2254 & 23.5 & 1972 & Europe \\ 

... &&&&&&&& \\
\hline
\end{tabular}
\end{center}
\end{table*}

\subsection{Demo 1: Encoding Number of Car Models Produced per Origin to Speech Pitch\protect\footnotemark} 

\begin{lstlisting}[language=json, numbers=none]
"encoding": {
    "time": {
        "field": "Origin",
        "type": "nominal"
    },
    "SpeechToneText":{
        "value": "Origin"
    },
    "SpeechTonePitch":{
        "aggregate": "count",
        "type": "quantitative",
        "scale": { "range": [0.75, 2.0] }
    }
}
\end{lstlisting}

We are interested in the number of car models produced per origin. Because this data is not directly provided in the dataset, we utilize the aggregation feature from \textit{Erie} to count the models. We encode the nominal attribute ``Origin'' into the time and ``SpeechToneText'' channels, such that the values of ``Origin'' are read out via speech in an order defined in the dataset (because it is nominal and does not have inherent order). We also encode the count of car models (per origin) into the pitch of speech with a range between 75\% and 200\% relative to the default pitch. When we play the speech sonification, the origins (e.g., ``Japan'', ``Europe'', etc.) are spoken linearly in time and pitched differently depending on the count of models from that origin. The country with the most number of car models sounds the highest in pitch.

\footnotetext[1]{Supplementary Material: \url{/Demo/Demo_1_Num-Car-Models-Per-Origin_Pitch.mp3}}
\footnotetext[2]{Supplementary Material: \url{/Demo/Demo_2_Num-Car-Models-Per-Year_Rate.mp3}}

\subsection{Demo 2: Encoding Numbers of Car Models Produced per Year to Speech Rate\protect\footnotemark}

\begin{lstlisting}[language=json, numbers=none]
"encoding": {
    "time": {
        "field": "Year",
        "type": "quantitative"
    }, 
    "SpeechToneText":{ 
        "value": "Year" 
    }, 
    "SpeechToneSpeed":{
        "aggregate": "count",
        "type": "quantitative",
        "scale": { "range": [1.2, 4.0] }
    }
}
\end{lstlisting}

We now want to know the trend of changes in the number of car models produced per year. We encode the quantitative attribute ``Year'' into the time and ``SpeechToneText'' channels, such that the values of ``Year'' are read out via speech sequentially. We also encode the count of car models each year into speech rate with a range between 120\% and 400\% relative to the default rate. When we play the speech sonification, the years (e.g., ``1971'', ``1972'', etc.) are spoken linearly in time with different rates depending on the count of models per year. The year spoken the fastest corresponds to the year with the highest number of car models produced.

\subsection{Demo 3: Encoding Fuel Efficiency to Time and Origin to Voice\protect\footnotemark}

\begin{lstlisting}[language=json, numbers=none]
"encoding": {
    "time": {
        "field": "Miles_per_Gallon",
        "type": "quantitative"
    },
    "SpeechToneText":{
        "value": "Name"
    },
    "SpeechToneVoice":{
        "field": "Origin",
        "type": "nominal",
        "scale": { "range": [65, 34, 0] }
    }
}
\end{lstlisting}

The goal of this example is to know what the most efficient car models were produced in 1982 measured by miles per gallon and which origin they were from. We encode the model names as the text content to be spoken out. We encode miles per gallon to time so that the data values are ordered by increasing fuel efficiency. We encode the three origins in voice and specify which voice we want to map. We map cars from Japan to a Japanese accent (voice ID 65 via Web Speech API on Chromium), Europe to a female voice (voice ID 34) and USA to the default male voice (voice ID 0). The sonification consists of car model names spoken in the increasing order of fuel efficiency with different voices depending on the origins.

\footnotetext[3]{Supplementary Material: \url{/Demo/Demo_3_Fuel-Efficiency_Origin_Voice.mp3}}

\section{Discussion}

In this section, we discuss the interrelation of speech-based encodings and the benefits and limitations of \textit{SpeechTone}. 

\subsection{Relationships between Speech-based Encodings}

Duration and Speed are not independent of each other and therefore cannot be mapped at the same time. 
Pitch can also play a significant role in our perception of a speaker’s gender. Higher pitches are often associated with female voices, while lower pitches are typically linked to male voices. This association can add a layer of complexity when considering the timbre of a voice in combination of the pitch channel. 
These intricate relationships between pitch, duration, speed and timbre make auditory encoding rich and complex, allowing \textit{SpeechTone} to leverage these relationships to create a multi-dimensional auditory experience.

\subsection{Benefits of \textit{SpeechTone}}

One of the key benefits of \textit{SpeechTone} is its ability to effectively condense information through speech encodings. The \textit{Erie} grammar originally uses speech to announce the mappings before the actual sonification to help users understand the sonification. By encoding data directly into speech attributes, one does not need to spend the time to listen to auditory cues to have contextual awareness of data labelling, potentially increasing the efficiency of sonification. 

Additionally, by introducing the \textit{SpeechTone} mark, we enable redundant encodings to attributes of both speech and musical tones. Such redundant encodings offer benefits such as reinforcing memory and combining numeric and auditory speech representations (similar to how FatFont~\cite{Nacenta_Hinrichs_Carpendale_2012} combines numeric and visual representations).

\subsection{Limitations and Future Work}

Further exploration is needed to examine the ranking of speech attributes in relation to human hearing abilities. This could involve investigating how different speech attributes are perceived and processed by the human auditory system and how these attributes can be effectively used in \textit{SpeechTone} to enhance data comprehension. While the three examples in Section~\ref{sec:demo} demonstrate the capabilities of \textit{SpeechTone}'s design, empirical validation is needed to confirm its effectiveness and representativeness across a broader range of data analysis tasks with screen-reader users. It would also be beneficial to investigate how \textit{SpeechTone} can be integrated with other sensory feedback mechanisms to provide a more comprehensive and immersive data analysis experience. 

Future work could additionally focus on improving the interactive capabilities of \textit{SpeechTone}, such as gazing, highlighting and zooming. Functionalities like real-time modification of encoding and search can also be beneficial.

\section{Conclusion}

We present a speech-based mark for sonification, \textit{SpeechTone}, an extension to the \textit{Erie} grammar, which introduces speech attributes as a new method for data mapping. This approach enhances the expressiveness and efficiency of sonification by directly encoding data into speech attributes such as pitch, speed, voice and text content. We demonstrate \textit{SpeechTone} through three examples. However, further exploration and empirical validation are needed to fully understand the potential of \textit{SpeechTone} in various data analysis tasks.

\section*{Supplementary Material}

The supplementary material provided with this paper includes several resources to help readers understand, replicate and extend the demonstrations discussed:

\begin{itemize}
    \item \path{Demo}: This directory contains three speech sonification samples, each corresponding to a specific demonstration described in Section~\ref{sec:demo}. 
    \item \path{SpeechTone-Erie-Demo.zip}: This is a sample project that creates the three speech demonstrations using \textit{SpeechTone}. 
    \item \path{SpeechTone-Erie-Source.zip}: This is the source code for the extended \textit{Erie} library, which incorporates \textit{SpeechTone}. 
\end{itemize}

\bibliographystyle{ACM-Reference-Format}
\bibliography{main}

\end{document}